\documentclass[twocolumn]{aa} 
\usepackage{amsmath}
\usepackage{graphicx}
\usepackage[utf8]{inputenc}
\usepackage{txfonts}
\usepackage{color}
\usepackage{cite}

\begin{document}

   \title{Missing large-angle correlations versus even-odd point-parity \\
          imbalance in the cosmic microwave background}

   \author{M.-A. Sanchis-Lozano$^1$, F. Melia$^2$\thanks{John Woodruff Simpson Fellow}, 
M. L\'opez-Corredoira$^3$ and N. Sanchis-Gual$^4$} 

   \offprints{M.-A. Sanchis-Lozano}
\titlerunning{Large-angle anomalies in the CMB}
\authorrunning{Sanchis-Lozano, Melia, L\'opez-Corredoira \& Sanchis-Gual}

\institute{$^1$Instituto de F\'{\i}sica Corpuscular (IFIC) and Departamento de F\'{\i}sica Te\'orica,
Centro Mixto Universitat de Val\`encia-CSIC, Dr. Moliner 50, E-46100 Burjassot, Spain; 
\email{Miguel.Angel.Sanchis@ific.uv.es} \\
$^2$Department of Physics, The Applied Math Program, and Department of Astronomy, 
The University of Arizona, Tucson, Arizona 85721, USA; \email{fmelia@email.arizona.edu} \\
$^3$Instituto de Astrof\'{\i}sica de Canarias, E-38205 La Laguna, Tenerife, Spain \\
Departamento de Astrofisica, Universidad de La Laguna, E-38206 La Laguna, Tenerife, Spain; 
\email{fuego.templado@gmail.com} \\
$^4$Departamento de Matem\'{a}tica da Universidade de Aveiro and Centre for Research and Development 
in Mathematics and Applications (CIDMA), Campus de Santiago, 3810-183 Aveiro, Portugal}

   \date{Received September 24, 2021}

 
  \abstract
   {The existence of a maximum correlation angle ($\theta_{\rm max} \gtrsim 60^\circ$)
in the two-point angular temperature correlations of cosmic microwave background (CMB)
radiation, measured by WMAP and {\it Planck}, stands in sharp contrast to 
the prediction of standard inflationary cosmology, in which the correlations
should extend across the full sky (i.e., $180^\circ$). The introduction of a hard lower
cutoff ($k_{\rm min}$) in the primordial power spectrum, however, leads naturally to the
existence of $\theta_{\rm max}$. Among other
cosmological anomalies detected in these data, an apparent dominance of odd-over-even
parity multipoles has been seen in the angular power spectrum of the CMB. This feature,
however, may simply be due to observational contamination in certain regions of the sky.}
   {In attempting to provide a more detailed assessment of whether this 
odd-over-even asymmetry is intrinsic to the CMB, we therefore proceed in this
paper, first, to examine whether this odd-even parity imbalance also manifests itself in
the angular correlation function and, second, to examine in detail the interplay between
the presence of $\theta_{\rm max}$ and this observed anomaly.} 
   {We employed several parity statistics and recalculated the angular correlation function
for different values of the cutoff $k_{\rm min}$ in order to optimize the fit to the different
\hbox{\it Planck} 2018 data.}
   {We find a phenomenological connection between these features in the data, concluding 
that both must be considered together in order to optimize the theoretical fit to the 
\hbox{\it Planck} 2018 data.}
   {This outcome is independent of whether the parity imbalance is intrinsic to the CMB, but
if it is, the odd-over-even asymmetry would clearly point to the emergence of new physics.}


   \maketitle
%

\section{Introduction} 
As is well known, observations of the temperature fluctuations in the cosmic microwave
background (CMB) radiation show that our Universe is quite uniform on scales much larger
than the apparent (or Hubble) horizon (Melia 2018) at the time of decoupling.
According to standard cosmology, this so-called horizon problem can be overcome by
assuming an inflationary phase lasting a tiny fraction of a second almost immediately
after the Big Bang (Starobinsky 1979; Kazanas 1980; Guth 1981; Linde 1982). Meanwhile,
quantum fluctuations in the underlying $\text{}$inflaton field $\phi$ (Mukhanov 2005)
would have grown and (somehow) classicalized to produce density perturbations that were also
stretched enormously by the accelerated expansion, eventually forming the seeds of today's
large-scale structure, including galaxies and clusters (Peebles 1980). Inflation has
also been invoked to explain why the Universe today appears to be spatially flat, if the
initial spatial curvature was indeed set arbitrarily (but see Melia 2022). If this initial 
condition were truly indeterminate, the Universe would have required an astonishing degree of 
fine-tuning at the time of the Big Bang to evolve into what we see today without the effects 
of inflation.

The most popular inflation models tend to adopt the slow-roll condition, positing that the
inflaton potential $V(\phi)$ changed very slowly during the phase of exponentiated expansion.
For these scenarios, it is convenient to measure the inflation time as a function of the
number of e-folds, $N_e$, in the expansion factor $a(t)$. The duration of inflation would then
be roughly equal to $N_eH_\phi^{-1}$, where $H_\phi\equiv \dot{a}/a$ was the Hubble parameter
at that time. In principle, the horizon and flatness problems might both be solved by
requiring an inflation corresponding to $N_e\gtrsim 60$, one of the more notable successes
of the inflation paradigm.

As we show below, however, the existence of a maximum correlation angle ($\theta_{\rm max}
\gtrsim 60^\circ$) observed in the CMB by all three major satellite missions,
COBE (Hinshaw et al. 1996), WMAP (Bennett et al. 2003), and {\it Planck} (Planck Collaboration 2018),
implies a smaller number of e-folds ($N_e\approx 55$), contrasting with most inflationary
scenarios (Melia \& L\'opez-Corredoira 2017; Liu \& Melia 2020; Melia et al. 2021). Moreover, 
this discrepancy does not stand in
isolation. Many other puzzles and anomalies contributing to an increasing level
of tension with standard cosmology ($\Lambda$CDM) have emerged in recent years as
the accuracy of the observations has improved (see, e.g., the recent review by 
Perivolaropoulos \& Skara 2021). For example, Schwarz et al. (2016)
included in their list of CMB anomalies an apparent alignment of the
lowest multipole moments with each other and with the motion and geometry of the Solar
System, a hemispherical power asymmetry, and an unexpectedly large cold spot in the
southern hemisphere. Di Valentino et al. (2021) argued against general
concordance by demonstrating that a combined analysis of the CMB angular power spectrum
obtained by {\it Planck} and the luminosity distance inferred simultaneously from type Ia
supernovae (SNe) excludes a flat universe and a cosmological constant at the $99\%$ confidence level.
A broader view of the general tension between the predictions of $\Lambda$CDM and
the observations may be found in L\'opez-Corredoira (2017). Whether some of these
anomalies have a common origin becomes of paramount importance to the fundamental basis
of our cosmological modeling.

In this paper we focus on two of the more recent discrepancies. The first is the
lack of large-angle correlations seen in the CMB data, which seems to suggest that the
primordial power spectrum, ${\mathcal{P}}(k)$, had a hard cutoff at a $k_{\rm min}$ distinctly
different from zero. This explanation for the angular-correlation anomaly has recently been
shown to also account self-consistently for the missing power at low $\ell$s in the angular
power spectrum (Melia et al. 2021). This feature in ${\mathcal{P}}(k)$ indicates the
time at which inflation could have started (Liu \& Melia 2020), hence setting an upper
limit to the possible number of e-folds by the time it ended. It is $k_{\rm min}$ that 
now appears to create an inconsistency between the number of e-folds required to solve 
the horizon problem and that corresponding to the measured fluctuation spectrum (Melia \&
L\'opez-Corredoira 2017).  The second is an apparent preference of the CMB data for an 
odd point-parity, first inferred from the analysis of the angular power spectrum (see, e.g.,
Kim et al. 2012; Schwarz et al. 2016). We seek to confirm whether this odd-even imbalance
is also present in the measured angular correlation function, and if it is, we attempt
to find a phenomenological connection between these two recently identified features in
the CMB fluctuation distribution.

We note, however, that an interpretation of the odd-even imbalance as being intrinsic
to the CMB anisotropies is not universally accepted. For example, Creswell \& Naselsky
(2021a) suggested that this asymmetry may be due to a contamination from a
few regions of the sky. We return to this viable possibility toward the end of
our discussion in \S~4. The work we carry out in this paper can therefore provide
a more quantitative assessment of the idea that the an odd-even imbalance may originate
within the CMB anisotropies for a more detailed comparison with the alternative scenario
in which it is primarily due to some observational contamination. In doing so, we
attempt to answer the questions whether (i)  either one or the other of these
characteristics is sufficient to account for the observed angular correlation function,
or if are both required; and (ii) if the latter is true, whether the reoptimized value of
$k_{\rm min}$ is different from that reported earlier (Melia \& L\'opez-Corredoira 2017;
Melia et al. 2021) and might mitigate the current level of tension between the latest 
{\it Planck} release and the predictions of standard inflationary cosmology.

In this paper, we do not, however, include the potentially useful
information concerning the polarization of the CMB radiation available from {\it Planck} for
several practical reasons. The E-mode polarization arising from the Thomson scattering of 
photons by free electrons is dominated by optically thin plasma on small spatial
scales. Therefore, these polarization effects cannot extend over large angular scales. In
addition, the subtraction of foreground contamination is more difficult to carry out for polarized 
CMB light based on the current {\it Planck} data because of the required complex 
multicomponent fitting. Future missions such as {\it Litebird} (Errard et al. 2016), 
{\it PICO} (Hanany et al. 2019) and {\it COrE} (Bouchet et al. 2011) will achieve more precise 
measurements at much higher sensitivity than is currently available with {\it Planck}, and 
should be able to answer the question of whether the large-angle anomalies seen 
in the temperature fluctuations are confirmed by the polarization maps.  

\section{Angular correlations in the CMB}
One of the main goals of analyzing the angular correlation function of the CMB is to
extract information regarding the different stages of the evolution of the Universe. Based
on very general grounds, small (large) angles between CMB photon
trajectories can be associated with small (large) length scales in the source plane (the opposite of
using energy scales). Furthermore, it offers us the possibility of analyzing an
important assumption in $\Lambda$CDM, that is, that the fluctuations are Gaussian and
statistically homogeneous and isotropic. As previously noted, the angular correlation
function of the CMB is already known to exhibit several statistical anomalies. In this
section, we focus on the lack of large-angle correlations, but we first recall features in the CMB that are of special interest to
this study.

\begin{figure}
\vskip 0.1in
\includegraphics[width=3.5in]{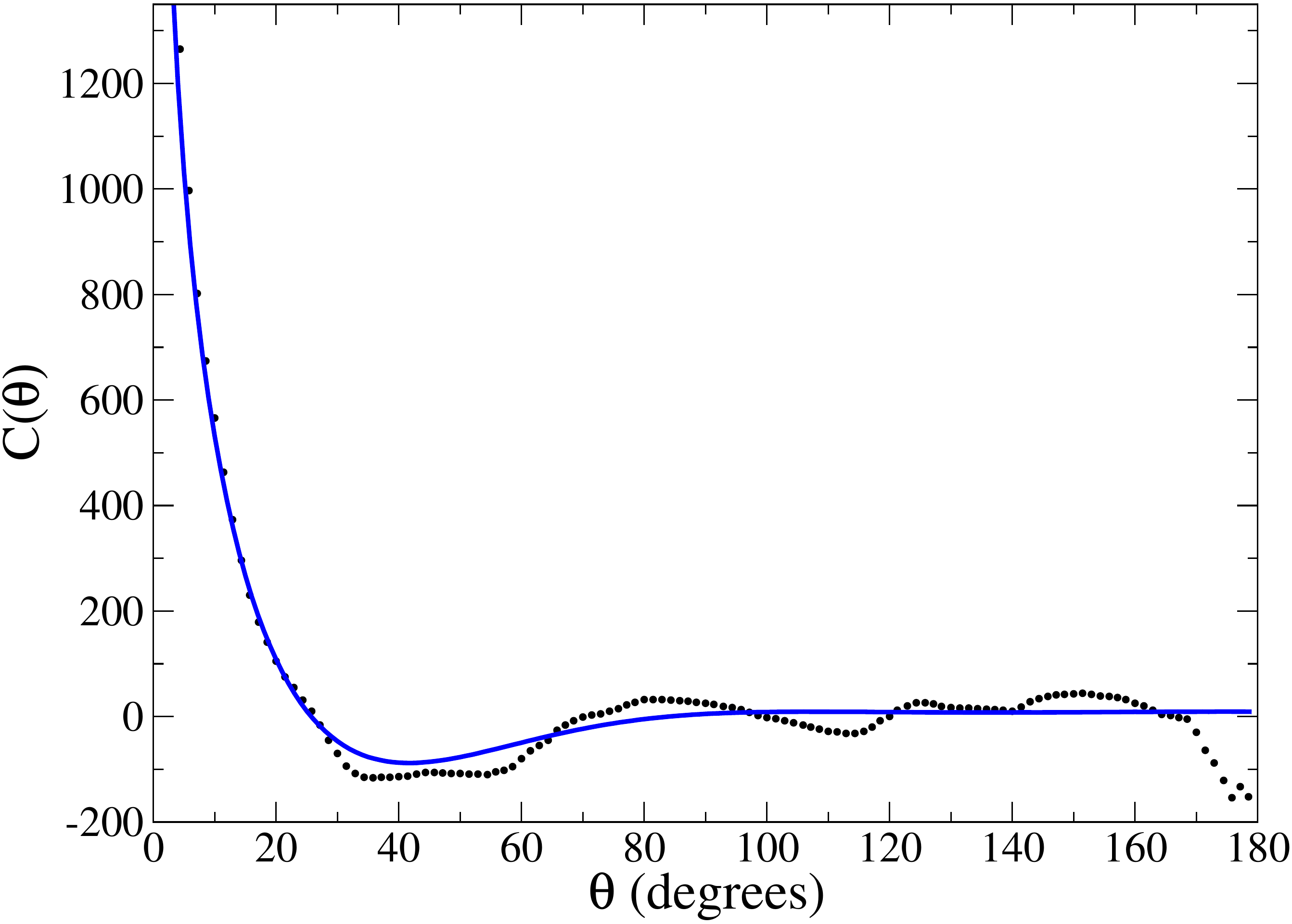}
\vskip 0.1in
\includegraphics[width=3.5in]{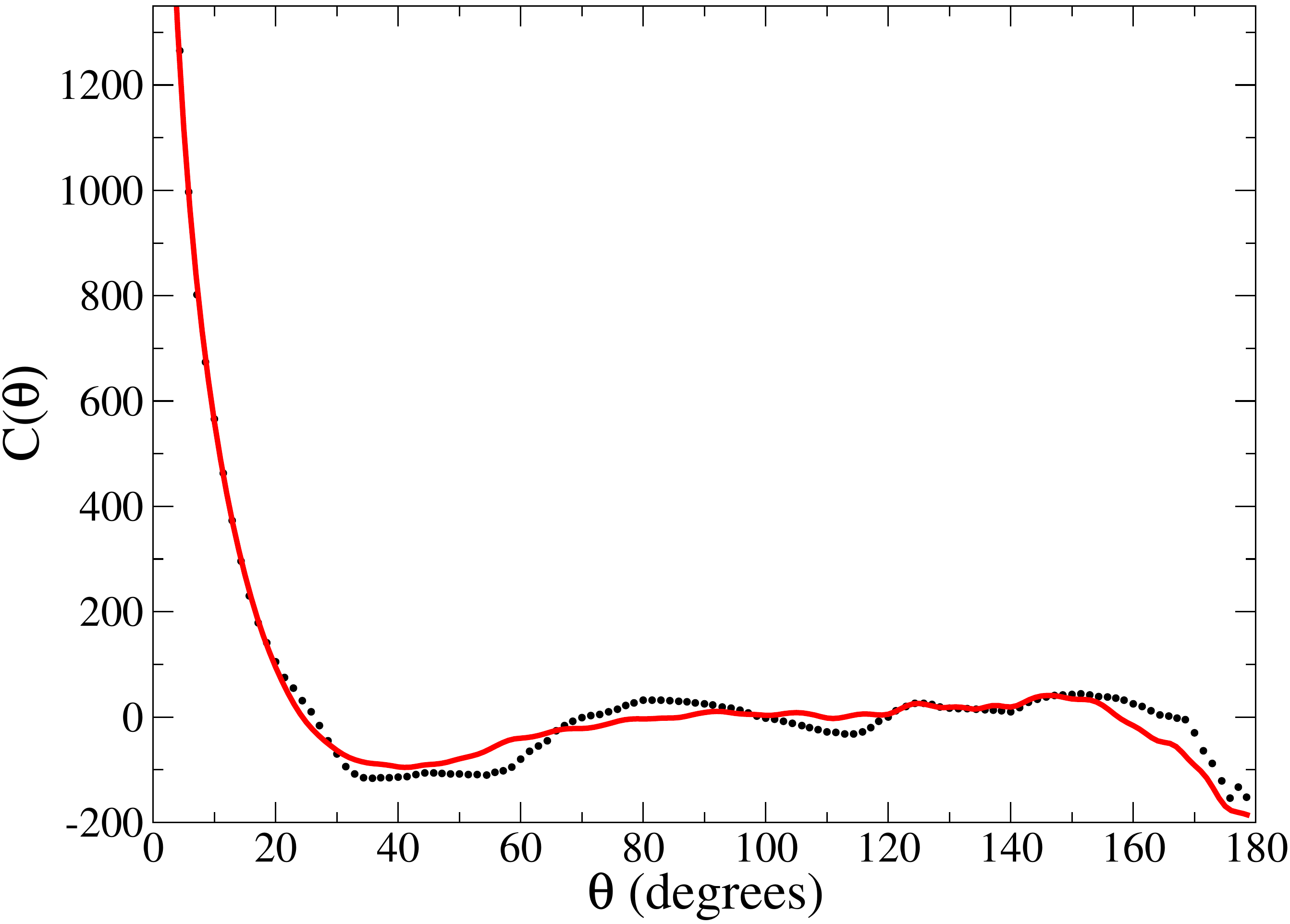}
\caption{Two-point correlation function $C(\theta)$ (solid curves) optimized to fit
the {\it Planck} 2018 data (black points) (Planck Collaboration 2018). (a) Top panel:
$u_{\rm min}=4.5$ and odd-even parity balance of multipoles (blue curve). (b) Bottom 
panel: $u_{\rm min}=4.5$ and odd-even parity dominance (red curve). See main text.}
\end{figure}

It is customary to distinguish primary anisotropies arising prior to decoupling,
from {\it secondary} anisotropies developed as the CMB photons propagate from the last
scattering surface to the observer. We do not distinguish between recombination,
last scattering surface, or freeze-out times, but approximate all of them as equal to 
a cosmic time $t_d \approx 3.8\times10^5$ years ($z \simeq 1100)$. For angles greater 
than a few degrees, the primary contributor to the former is the Sachs-Wolfe (SW) effect 
(Sachs \& Wolfe 1967), representing fluctuations in the metric leading to temperature
anisotropies via perturbations of the gravitational potential at the time of decoupling.
There are two Sachs-Wolfe influences: the above-mentioned, nonintegrated Sachs-Wolfe
effect, and the integrated Sachs-Wolfe effect (ISW). The latter is sometimes also further
subdivided into an early ISW (taking place just after decoupling; for convenience, this
is often just included in the SW), and a late ISW, arising while the CMB photons propagated
through the expanding medium. The ISW contributes non-negligibly to the CMB anisotropies
only if the universal expansion is at least partially driven by something other than
purely nonrelativistic matter. In the standard model, dark energy---possibly in the
form of a cosmological constant---started influencing the expansion at $z\sim 0.5$,
corresponding to a cosmic time $\sim 10$ Gyr. The ISW has the effect of mainly raising
the Sachs plateau at low multipoles. However, the detection of this ISW
due to dark energy is not fully confirmed yet. The putative detections may simply
be noise with underestimated error bars (L\'opez-Corredoira et al. 2010; Dong et al. 2021).

A way to address the angular dependence and anisotropies of the CMB is through its primary
power spectrum, originally defined by the Fourier transform of the primordial fluctuation
spectrum, usually parameterized as
\begin{equation}
P(k)=A\ \biggl[\frac{k}{k_0}\biggr]^{n_s-1}\;,\label{eq:primordial}
\end{equation}
where $n_s$ is the scalar spectral index. The spectrum would be perfectly scale free
(with $n_s=1$) if the Hubble parameter $H_\phi$ were strictly constant during inflation.
In typical slow-roll inflationary models, however, this is only approximately true,
and $H_\phi$ evolves slowly, which produces a slight deviation of the spectral index
from one. The observations show that in fact $n_s=0.9649 \pm 0.0042$ (Planck Collaboration 2018),
adding some observational support for a slow-roll potential, $V(\phi)$.

\subsection{Two-point angular correlation function}
The anisotropies in the CMB are very small, of about one part in $10^5$, but they carry
a wealth of information pertaining to the possible influence of $V(\phi)$ and the
subsequent evolution of the Universe after reheating. A very powerful probe of these
fluctuations is the two-point angular correlation function, defined as the ensemble product
of the temperature differences with respect to the average temperature, from two directions
in the sky defined by unitary vectors $\vec{n}_1$ and $\vec{n}_2$,
\begin{equation}\label{eq:CTT}
C(\theta)=\langle T(\vec{n}_1)T(\vec{n}_2)\rangle\;.
\end{equation}
The angle $\theta \in [0,\pi]$ is defined by the scalar product $\vec{n}_1 \cdot \vec{n}_2$.

One typically expands $C(\theta)$ in terms of Legendre polynomials (assuming azimuthal
symmetry)
\begin{equation}\label{eq:CCC}
C(\theta)=\frac{1}{4\pi}\ \sum_{\ell=2}^{\infty}\ (2\ell+1)C_{\ell}P_{\ell}(\cos{\theta})\;,
\end{equation}
where the $C_{\ell}$ coefficients encode the information with cosmological significance from
the sky. The sum starts at $\ell=2$ and ends at a given $\ell_{\rm max}$, dictated by the
resolution of the data. The first two terms are excluded because (i) the monopole ($\ell=0$)
is simply the average temperature over the whole sky and plays no role in the
correlations, other than a global scale shift; and (ii) the dipole ($\ell=1$) is
greatly affected by Earth's motion, creating an anisotropy that dominates the intrinsic
cosmological dipole signal.

\subsection{Maximum angle in two-point angular correlations}
Large-angle correlations in the CMB provide information about the earliest stages of the
primitive Universe, well before recombination and the subsequent formation of cosmic
structure. In this context, it may be useful to point out an interesting analogy with the
angular correlations among the final-state particles in heavy-ion or proton-proton
collisions at the Large Hadron Collider, and the early formation of nonconventional
matter, such as a quark-gluon plasma or hidden valley particles (Sanchis-Lozano et al. 2020).
In Figure~(1a) we plot the observed angular correlation function (black dots) measured by 
{\it Planck} (Planck Collaboration 2018), compared with fitted (blue and red) curves to be discussed
below. Above $\simeq 60^\circ$ , the correlations drop to near zero, except for
a downward tail at $\sim 180^\circ$, which we also examine in more detail below. This
shape of $C(\theta)$, particularly its suppression at large angles, was unexpected in standard 
cosmology, given that inflation was supposed to begin early enough (with $k_{\rm min}\rightarrow 0$) 
to provide the required number of e-folds to solve the horizon and flatness problems, thereby 
providing coverage across the full sky (see, e.g., Melia 2014).

It is worth mentioning at this point that these expectations on the value of
$k_{\rm min}$ and correlations at all angles are primarily based on the correctness
of the standard model. Large-angle correlations are not necessarily expected
in all cosmological models, however. For example, in the alternative cosmology 
known as the $R_{\rm h}=ct$ universe, which does not have an inflationary epoch, 
the expansion factor is linear in time and the maximum angle corresponds to the size 
of the apparent horizon (Melia 2018) at decoupling,
\begin{equation}
\theta_{\rm max}\ \simeq\  \frac{2\pi}{\ln{(t_0/t_d)}}\quad({\rm in\; radians})\;,\label{eq:thmax}
\end{equation}
where $t_d$ and $t_0$ denote the decoupling and present cosmic times, respectively.
The $R_{\rm h}=ct$ universe is an FLRW cosmology in which the equation of state is 
constrained by the zero active mass condition in general relativity, that is, $\rho+3p=0$. The 
Raychaudhuri equation clearly shows that $\ddot{a}=0$ in that case, which leads to a Universe 
expanding at a constant rate (Melia 2013a). This Universe has no horizon problem, 
and spatial flatness is ensured because the total energy density is zero. It therefore 
has no need for inflation. In such a universe, the Hubble parameter is always exactly 
equal to the inverse of the age of the Universe, and the Hubble radius satisfies 
$R_{\rm h}=ct$ at all times, hence the eponymous origin of its name. Setting 
$t_d=3.8 \times 10^5$ years and $t_0=13.8$ Gyr, we obtain $\theta_{\rm max} \sim 40^\circ$.  
Thus, if inflation were to fail to adequately explain the existence of such a maximum correlation
angle, it might be considered evidence supporting a noninflationary model, such as $R_{\rm h}=ct$.

We generalize the expression in Equation~(\ref{eq:thmax}) by writing it in terms of
the maximum fluctuation size $\lambda_{\rm max}(t_d)$ at decoupling time $t_d$
and the proper distance $R_d\equiv a(t_0)r_d$ (with $r_d$ the comoving distance)
from us to the last scattering surface,
\begin{equation}\label{eq:thetamax}
\theta_{\rm max}= 2\ \tan^{-1} \biggl[ \frac{\lambda_{\rm max}}{R_d} \biggr]\;.
\end{equation}
Because we study the formation of primordial physical quantities, the apparent
horizon (equal to the Hubble radius in this case) should set the basic length scale at stake
(Melia 2013b, 2020b).  In particular, the maximum fluctuation size can be estimated
as $\lambda_{\rm max} = \alpha\ 2\pi R_{\rm h}$, where $\alpha \lesssim 1$ denotes a coefficient
dependent on the cosmological model. For instance, $\alpha=1$ for de Sitter space and $\alpha 
\simeq 0.5$ for $\Lambda$CDM (Melia 2013b, 2018). This scale changes as the expansion
factor $a(t)$ grows, starting with the assumed slow-roll inflation, followed by radiation
and then matter-dominated evolution, before reaching decoupling, where the CMB radiation
was released. Nevertheless, knowledge of $R_{\rm h}$ at decoupling is sufficient to estimate
$\lambda_{\rm max}$ associated with the largest fluctuation we can see in the CMB anisotropies
today.

\begin{figure}
\vskip 0.1in
\includegraphics[width=3.5in]{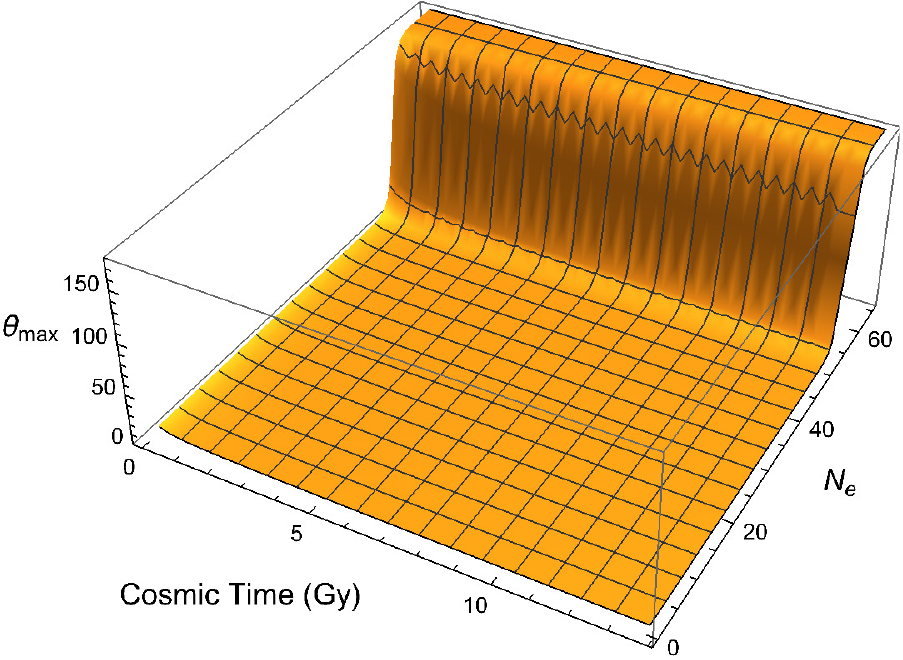}
\vskip 0.2in
\includegraphics[width=3.5in]{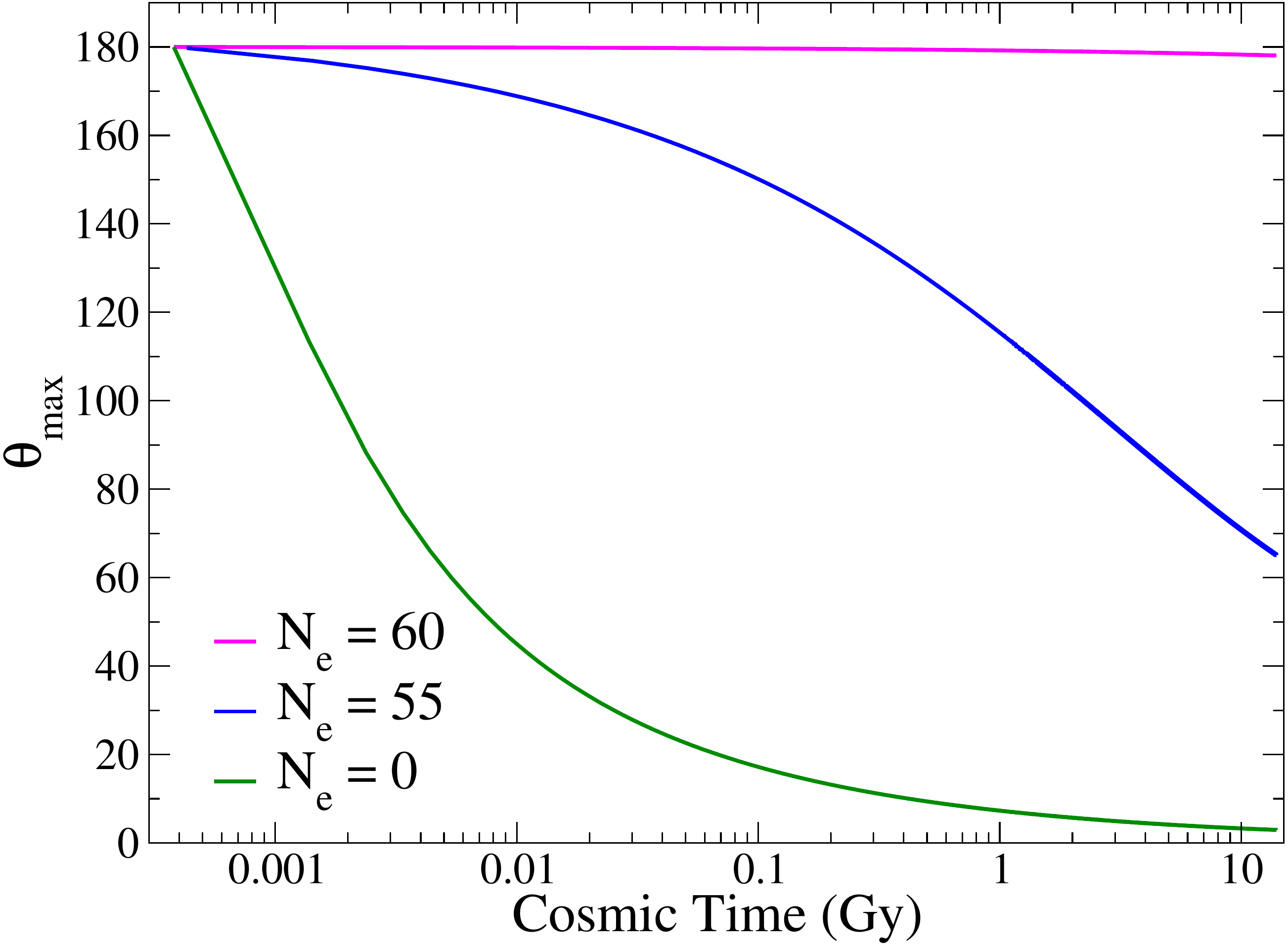}
\caption{Maximum correlation angle due to inflation. 
Top panel (a): 3D plot of the maximum correlation angle as a function of
cosmic time and the number of e-folds. Bottom panel (b): Maximum correlation angle
evolution with cosmic time for different $N_e$ values, corresponding to vertical
sections in the 3D plot. We highlight that the plateau for $N_e$ is smaller than about
50, implying a large inhomogeneity of the CMB along cosmic time.}
\end{figure}

On the other hand, to compute the proper distance $R_d$ between us and the last scattering
surface, we must know the expansion history from decoupling to today. An excellent approximation
for $a(t)$ during this time may be written
\begin{equation}
a(t) \sim \sinh^{2/3}(\tilde{H}\ t)\;,
\end{equation}
where $\tilde{H} \equiv (3/2)\sqrt{\Omega_{\Lambda}}H_0$, with $\Omega_{\Lambda}$ and
$H_0$ denoting the normalized dark matter density and Hubble constant today,
respectively, and $t$ the cosmic time since the Big Bang (see Appendix A).

In Figure~2 we show the maximum correlation angle obtained in our study under different
assumptions concerning the inflationary epoch. Figure~2(a) shows a 3D rendition of the
maximum correlation angle as a function of cosmic time and the number of e-folds
($N_e$). The correlation angle remains confined to just a few degrees
for any number of e-folds $N_e \lesssim 40$. For $N_e \geq 62$, on the other hand,
the curve quickly reaches $180^\circ$ along the whole cosmic time until now. For the
sake of clarity, the maximum correlation angle evolution with cosmic time is shown
in Figure~2(b) for different $N_e$ values, corresponding to vertical cuts in the
3D plot (Fig.~2a) (see Appendix A for the mathematical details).

The late-time ISW effect yielding additional secondary anisotropies should tend to broaden
the angular correlations as lower-$\ell$ multipoles are enhanced (due to a raising of
the Sachs-plateau). Therefore, the number of e-folds required to comply with the
observed maximum value of $\theta_{\rm max}$ should decrease even further once the ISW is
taken into account. The {\it Planck} 2018 data therefore suggest that $N_e \lesssim 55$,
creating even greater tension with the conventional slow-roll inflationary
scenario. Liu \& Melia (2020) provided more details concerning the difficulties
faced by the slow-roll paradigm to simultaneously solve the horizon problem and missing
correlations at large angles.

Addressing these constraints from the latest CMB data would require a more complicated
inflationary process than is usually conjectured. As we show below, this conclusion
goes in the same direction as the need for a cutoff in the power spectrum in order to
correctly reproduce the whole angular correlation function of the CMB.

\subsection{Low cutoff in the CMB power spectrum}
$\Lambda$CDM predicts an angular correlation curve that crosses the zero-axis twice
and extends over the whole $180^\circ$ range of poloidal angles (see, e.g., Fig.~1 in
Melia 2014). This result is manifestly inconsistent with observational evidence, which 
shows a maximum correlation angle of $\geq 60^o$, as already discussed in the previous 
section.

In order to mitigate this tension, Melia \& L\'opez-Corredoira (2017) 
introduced a cutoff to the primordial power spectrum, representing a lower limit to 
the integral
\begin{equation}\label{eq:Cell}
C_{\ell}\ =\ N\ \int_{k_{\rm min}}^{\infty}dk\ k^{n_s-1}\ j_{\ell}^2(k\ r_{d})\;,
\end{equation}
where the normalization constant $N$ and the minimum mode wavenumber $k_{\rm min}$
are optimized using a global fit to the whole observed angular correlation function.
Although this procedure represents a phenomenological introduction of the cutoff
$k_{\rm min}$, this truncation has some theoretical justification in that it
represents the first quantum fluctuation to either (i) have crossed the Hubble horizon
once inflation started, or (ii) have emerged out of the Planck domain if inflation never 
happened (see Melia 2019; Liu \& Melia 2020). The need for a $k_{\rm min}$ 
may also be related to the infrared regularization of the inflaton field commutator, 
although in standard cosmology, it should then likely be much smaller than its value 
(i.e., $\simeq 3\times 10^{-4}$ Mpc$^{-1}$, corresponding to $u_{\rm min}=4.34$) 
optimized using the {\it Planck} 2018 data (Liu \& Melia 2020).

In the computation of the $C_{\ell}$ coefficients, only the SW effect is taken into
account, ignoring other effects such as the baryon acoustic oscillations (BAO),
whose influence extends primarily over smaller angles ($\theta \lesssim 5^\circ$)
and hence only the very large-$\ell$ multipoles (certainly $>100$).
Changing the integration variable from $k$ to $u\equiv kr_d$ in Equation~(\ref{eq:Cell})
and setting $n_s=1$ for simplicity, we obtain 
\begin{equation}\label{eq:Cumin}
C_{\ell}\ =\ N\ \int_{u_{\rm min}}^{\infty}du\ \frac{j_{\ell}^{\,2}(u)}{u}\;.
\end{equation}
We  point out that only those $C_{\ell}$
coefficients with $\ell \lesssim 20$ are actually affected by the existence of the
cutoff $u_{\rm min}$ in the above integral.

In a previous computation of these coefficients using the {\it Planck} 2103 dataset, 
Melia \& L\'opez-Corredoira (2017) found that the best fit to the angular correlation 
function is obtained with $u_{\rm min}=4.34\pm 0.50$, which translates into a minimum wavenumber 
$k_{\rm min}=4.34/r(t_d)$ (see also the similar limit placed on $k_{\rm min}$ by an analogous 
study of the angular power spectrum itself; Melia et al. 2021). In the present paper, we
repeated this analysis using the more recent {\it Planck} 2018 dataset, obtaining $u_{\rm min}=
4.5\pm 0.5$, and $k_{\rm min}=4.5/r(t_d)$, which is compatible with the previous results from {\it Planck} 
2013. We provide more details about the statistical analysis yielding this result
below.

Very importantly, an almost zero correlation plateau above the maximum angle
($\theta_{\rm max}\approx 60^o$) can be obtained by setting a lower cutoff to
the integration variable $u$, corresponding to a lower cutoff in the power spectrum.
Mathematically, this result can be understood as a delicate balance between even
and odd multipole contributions to $C(\theta)$. We return to this crucial
point in the next section.

\section{Odd versus even point-parity in the CMB}
Among the other anomalies observed in the CMB, an odd-even parity violation may indicate
a nontrivial topology of the Universe, unexpected physics at the pre- or inflationary
epochs, or some unsolved systematic errors in the data reduction. In the following, we focus on the apparent odd-dominance of the CMB fluctuations, that is, the fact that
the weight of odd multipoles in either the power spectrum or the two-point angular correlation
function is larger than the corresponding weight of the even multipoles. This imbalance is
commonly referred to as a point-parity asymmetry of the CMB, and we address two
statistics below that are widely employed to analyze it.

\subsection{Odd-even parity statistics}
We employ the parity statistic (Panda et al. 2020)
\begin{equation}
P(\ell_{\rm max})=\frac{P^+(\ell_{\rm max})}{P^-(\ell_{\rm max})}\;,
\end{equation}
where
\begin{equation}
P^{\pm}(\ell_{\rm max})=\sum_{\ell=2}^{\ell_{\rm max}}\ 
\gamma_{\ell}^{\pm}\ \frac{\ell(\ell+1)}{2 \pi}\ C_{\ell}\;,
\end{equation}
with the projectors defined as $\gamma_{\ell}^+=\cos^2{(\ell\pi/2)}$ and
$\gamma_{\ell}^-=\sin^2{(\ell\pi/2)}$. Assuming that $\ell(\ell+1)C_{\ell}$ is approximately 
constant at low $\ell$, $P^{\pm}$ can clearly be considered as a measurement of the degree
of parity asymmetry: below unity, it implies odd-parity dominance, and vice versa. Any deviation
of this statistic from unity points to an odd-even parity imbalance.

\begin{figure}
\vskip 0.1in 
\includegraphics[width=3.5in]{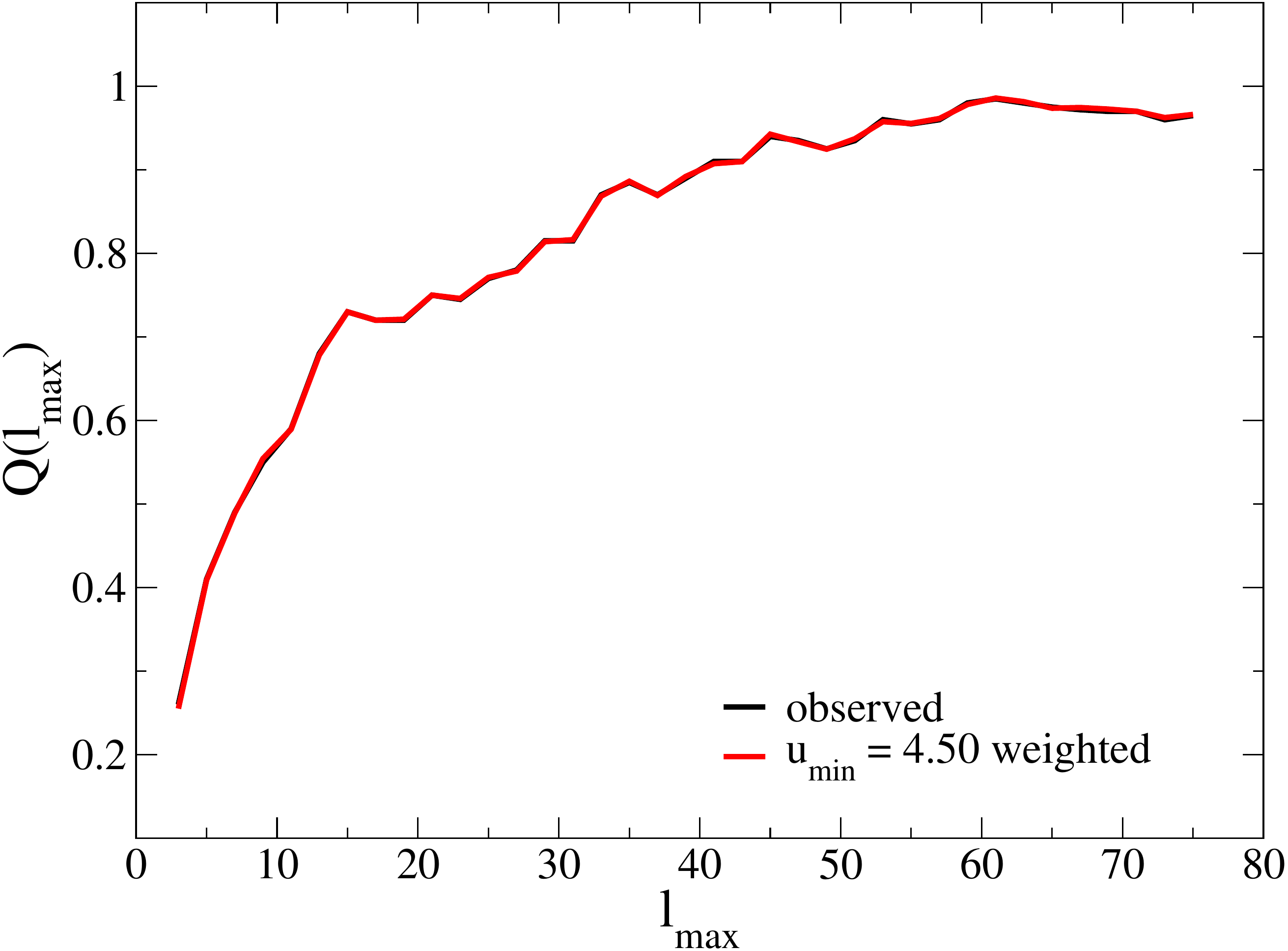}
\vskip 0.2in
\includegraphics[width=3.5in]{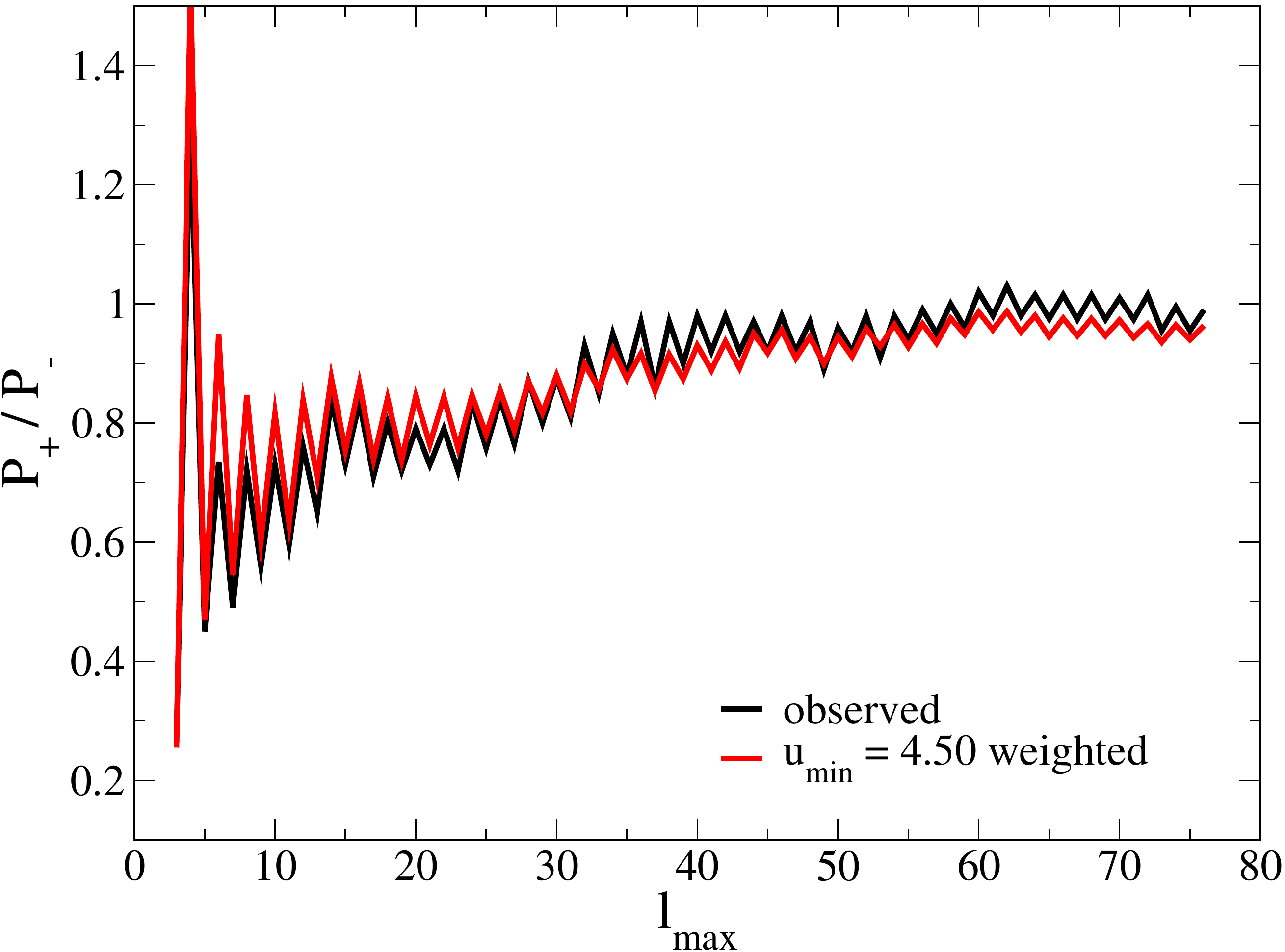}
\caption{Statistics used to describe the point-parity asymmetry of the CMB.
(a) Top panel: $Q(\ell_{\rm max}^{\,\rm odd})$ statistic as a function of
$\ell_{\rm max}$ for $u_{\rm min}=4.50$ (red), compared to the {\it Planck} 2018 data. The overlap
is visually almost perfect.
(b) Bottom panel: Same for the $P(\ell_{\rm max})$ statistic using the values of
$C_{\ell}$ determined from the matching of the $Q(\ell_{\rm max}^{\,\rm odd})$ statistic
with the data shown on the left. The reduced $\chi^2$ for this fit is $\chi^2_{\rm dof}\approx 0.99$.}
\end{figure}

A second statistic useful for checking a point-parity imbalance may be defined as the average
ratio of the power in adjacent odd and even multipoles up to a given $\ell$ value
(Aluri \& Jain 2012; Panda et al. 2020),
\begin{equation}
Q(\ell_{\rm max}^{\,\rm odd})=\frac{2}{\ell_{\rm max}^{\,\rm odd}-1}\ 
\sum _{\ell=3}^{\ell_{\rm max}^{\,\rm odd}}\ \frac{D_{\ell}-1}{D_{\ell}}\;,
\end{equation}
where $\ell_{\rm max}^{\,\rm odd}$ is the maximum odd multipole up to which the statistic
is computed, and $D_\ell\equiv \ell(\ell+1)C_\ell/\pi$. In contrast to
$P(\ell_{\rm max})$, the new statistic, $Q(\ell_{\rm max}^{\,\rm odd})$, ensures that
there are always the same number of odd and even powers along the whole considered
multipole range, so no sawtooth oscillations are present. As for $P(\ell_{\rm max})$,
this statistic is also expected to fluctuate about the value of one at low $\ell$s.

In case of Gaussian fluctuations, the angular power spectrum and the angular correlation
function contain the same information concerning the angular distribution of temperature
in the CMB. Nevertheless, although the angular power spectrum covers all of the
$\ell$ dependence, it emphasizes large-$\ell$ values, such that most of the information
at $\theta \gtrsim 10^\circ$ is squeezed into a very narrow interval, making it difficult
to pick out any disagreement between theory and observation for the low multipoles.
Conversely, the angular correlation function covers the fluctuation distribution more
evenly over all angles, thereby making it relatively easier to study the large-angular
region, corresponding to lower multipoles. Ultimately, both approaches should be mutually
consistent as well as complementary for the extraction of useful information.

With this goal in mind, we require the statistic $Q(\ell_{\rm max}^{\,\rm odd})$ (shown
in red) to match (up to a given accuracy) the data (shown in black) in Figure~3(a). We do
this by heuristically tuning the weights
\begin{equation}\label{eq:ql}
q(\ell_{\rm even})=\frac{C_{\ell_{\rm even}}}{C_{\ell_{\rm even}+1}}
\end{equation}
to optimize the fit. Then, using the above ratios, we fit the two-point angular correlation
function keeping two parameters free, namely, the normalization, $N$, and the cutoff
$k_{\rm min}$ (i.e., $u_{\rm min}$). As noted earlier, the latter was already constrained
to the interval $u_{\rm min} \in (4.34 \pm 0.50)$ in Melia \& L\'opez-Corredoira (2017) using
the old {\it Planck} 2013 data and without any consideration of a possible odd-even parity 
imbalance. We have carried out this analysis again for the latest {\it Planck} 2018 data at 
small, middle, and now large angles, incorporating the odd-parity dominance.

\begin{figure}
\vskip 0.1in
\includegraphics[width=3.5in]{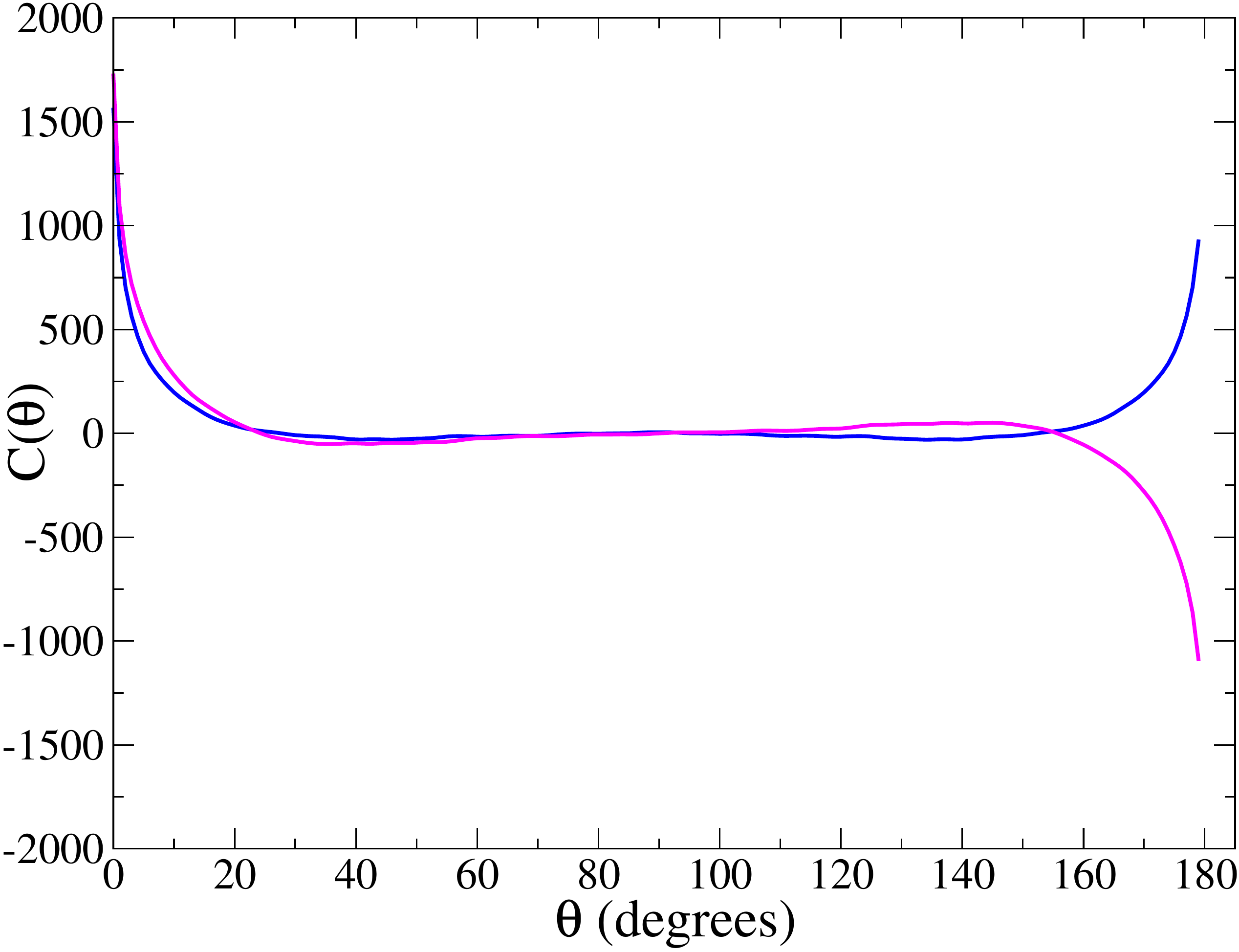}
\vskip 0.2in
\includegraphics[width=3.5in]{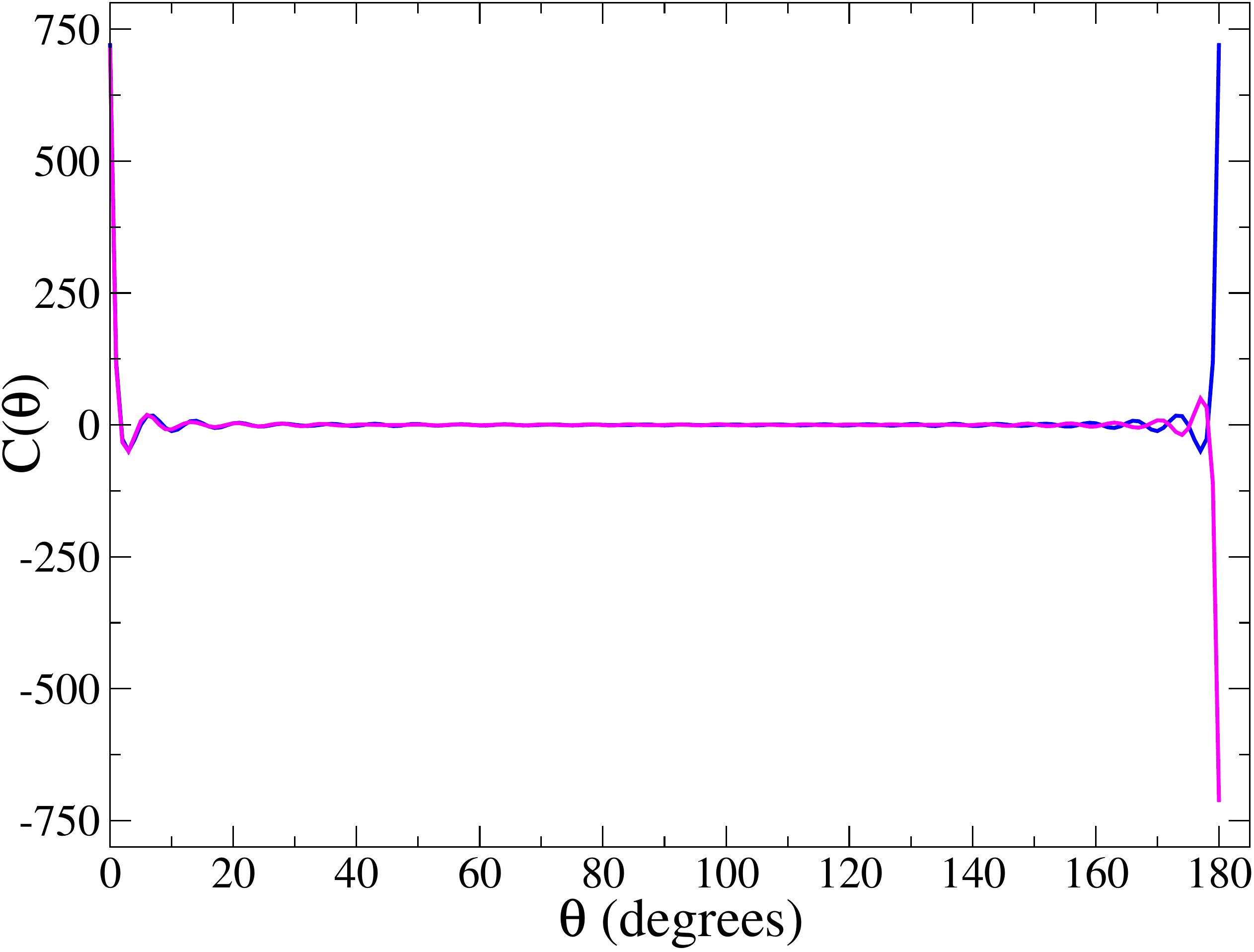}
\vskip 0.2in
\caption{Contributions from the even (blue) and odd (magenta) multipoles to the
two-point angular correlation function $C(\theta)$ in Equation~(\ref{eq:CCw})
for (a) $\ell \in [2,401]$ (left panel) and (b) $\ell \in [40,401]$ (right panel).
The delicate balance between odd and even polynomials is clear even at high $\ell$ for
large angles. Selecting a different weighting of each contribution has a strong influence
on $C(\theta)$ at $\theta\simeq 180^\circ$.}
\end{figure}

To incorporate these weights into the angular correlation function, we modified
Equation~(\ref{eq:CCC}) to read
\begin{equation}\label{eq:CCw}
C(\theta)=\frac{1}{4\pi}\ \sum_{\ell=2}^{\infty}\ (2\ell+1)\ w_{\ell}\ 
C_{\ell}P_{\ell}(\cos{\theta})\;,
\end{equation}
with
\begin{equation}
w_{\ell_{\rm even}}=\frac{q(\ell_{\rm even})}{q(\ell_{\rm even})+1},\ \ \ 
w_{\ell_{\rm even}+1}=1-w_{\ell_{\rm even}}\;,
\end{equation}
where $\ell_{\rm even}$ denotes even values of $\ell$ within the selected interval.
The weights $w_{\ell}$ for the coefficients $C_{\ell}$, introduced ad hoc to model 
the odd-even imbalance, should not be confused with the window factors.  Obviously, 
by requiring $q(\ell_{\rm even}) \equiv 1$ $\forall\; \ell_{\rm even}$, we obtain 
$w_{\ell_{\rm even}}=w_{\ell_{\rm even}+1}=1/2$, thereby restoring the odd-even parity
balance.

In Figure~3(a) we show the function $Q(\ell_{\rm max})$ for the {\it Planck} 2018 data
(black) with the fit (red) corresponding to the set of parameters that provides the
minimum reduced $\chi^2_{\rm dof}$ for the angular two-point correlations, keeping the cutoff
we introduced to improve the fit (see Table~1). The matching of the observed and theoretical values
achieved by tuning the ratios $q(\ell)$ (and their respective weighting factors,
$w_{\ell}$) that modify the $C_{\ell}$ coefficients is excellent. In addition,
the set of weights optimized to find this match for $Q(\ell_{\rm max})$ also produces
an excellent fit in the $P(\ell_{\rm max})$ plot, corresponding to a reduced $\chi^2_{\rm dof}$
near unity, as seen in Figure~3(b) and Table~1.

Figure~1(b) shows the plot of $C(\theta)$ over the whole range of angles with
$u_{\rm min}=4.5$, although this time, it incorporates the odd-dominance via the coefficients
$C_{\ell}$, modified as explained above. The fit now also matches the downward tail at large
angles. To understand what is happening, consider the separate plots in Figure~4(a) that 
show the odd and even multipole contributions to $C(\theta)$.  Each (odd
and even) contribution behaves quite distinctly due to the intrinsic parity properties of
the Legendre polynomials about $\theta=90^\circ$.  Consequently, once summed to produce
the full $C(\theta)$, these two contributions produce constructive interference at small
angles (i.e., $\theta< 90^\circ$) and destructive interference at large angles ($\theta > 
90^o$).

\begin{table*}[hbt]
\setlength{\tabcolsep}{0.5pc}
\caption{$\chi^2_{\rm dof}$ values for different fits to the two-point correlation function $C(\theta)$
and the parity statistic $P(\ell_{max})$, inferred from the {\it Planck} data. PB: parity balance; OD: odd dominance}
\label{tab:Tab1}

\begin{center}
\begin{tabular}{ccccc}
  \hline 
&&&& \\
Curve/parameters: &  $u_{min}=0$+PB   &  $u_{min}=0$+OD  &
$u_{min}=4.5$+PB & $u_{min}=4.5$+OD \\
&&&& \\
\hline
$C(\theta)$ & $6.1$  & $3.8$ & $0.80$ & $0.70$ \\
$P(\ell_{max})$ & $30$ & $5.3$ & $18.0$ & $0.99$ \\
\hline

\end{tabular}
\end{center}
\end{table*}

One of our most important conclusions from this work is that in order to produce the
optimum fit of the {\it Planck} 2018 data with the lowest $\chi^2_{\rm dof}$ in Figure~1(b),
we must include in our theoretical $C(\theta)$ both the cutoff $u_{\rm min}$ in
Equation~(\ref{eq:Cumin}) and an odd-parity dominance via the weighting factors in
Equation~(\ref{eq:CCw}). Otherwise, our fit to the observed data points for either the
angular correlation function or the parity statistics worsens considerably, yielding a
$\chi^2_{\rm dof}$ much larger than one. Using both the cutoff ($k_{\rm min} \neq 0$)
and an odd-even parity imbalance simultaneously produces the best fit for $C(\theta)$,
$Q(\ell_{\rm max})$ and $P(\ell_{\rm max})$ (see Table~1), however. 

As noted earlier, we find from our study an optimized cutoff
$u_{\rm min} \simeq 4.5 \pm 0.5$, in which the mean value and $\sigma=0.5$ error were obtained
using a Monte Carlo analysis to sample the variation of $C(\theta)$ within the measurement errors. 
Since the  $C(\theta)$ points are highly correlated, we circumvented this problem using
a Monte Carlo procedure as described in Melia \& L\'opez-Corredoira (2017). Starting from 100
mock CMB catalogs, we computed the two-point correlation function $C_i(\theta)$, ($i=1,100$) 
in each case. From these, we obtained $\Delta C_i(\theta)=C_i(\theta)-C_0(\theta)$, where 
$C_0(\theta)$ is the angular correlation function in standard cosmology. Then we calculated  
$u_{{\rm min} ,i}$ for $C(\theta) = C_{Planck}(\theta) + \Delta C_i(\theta)$ for each realization 
$i$. Next, from the resulting roughly Gaussian distribution of $u_{{\rm min} ,i}$, we determined
its average value and r.m.s., yielding $u_{\rm min} \simeq 4.5 \pm 0.5$, whose corresponding 
$\chi^2_{\rm dof}$ distribution from the $C(\theta)$ fits varies smoothly around its minimum 
at 4.5. This $u_{min}$ range is slightly larger than but compatible with the interval obtained 
from the {\it Planck} 2013 dataset: $u_{\rm min}=4.34\pm 0.50$.

A quantitative comparison of the various fits we have explored is provided in Table 1, which
lists the $\chi^2_{\rm dof}$ values for different choices of parameters, including the odd-even parity 
relative weights. A remark is in order: the improvement of the $C(\theta)$ fit, once $u_{min}=4.5$ is 
fixed, is somewhat modest when passing from parity balance to parity breaking. This is easy to 
understand because the main effect of imposing odd-dominance takes place at large angles,
where the observational uncertainties are quite large and its impact on the total $\chi^2_{\rm dof}$ 
is thus relatively small. On the other hand, the agreement between the theoretical prediction and 
the parity statistic $P(\ell_ {max})$ improves dramatically, as expected, given the observed 
odd-dominance, as seen in Figure~3(a).

We stress that the main benefit of using a $k_{min} \neq 0$ {\em together with} parity breaking
is not so much the improvement of the fit to the two-point correlations (with a focus on the tail at 
large angles) at the cost of increasing the number of degrees of freedom, but the fact that this 
approach can simultaneously resolve several apparently disconnected anomalies in the CMB. The end 
result is an excellent fit of $C(\theta)$ at small, medium, and large angles (reproducing the 
existence of a $\theta_{max} \gtrsim 60^\circ$ ) with a p-value $\simeq 0.95$, and the implementation 
of an odd-parity dominance seen also in the angular power spectrum. This outcome constitutes the 
principal result of our analysis in this paper.

\subsection{Relevance of the angular correlations at $\theta \simeq 180^\circ$}
Large-angle correlations are commonly associated with low multipoles, typically
$\ell \in [2,20]$. We have already established that the introduction of a cutoff
$k_{\rm min}$ mainly affects multipoles with $\ell \lesssim 20$, and
significantly alters the shape of the $C(\theta)$ curve at $\theta \gtrsim 60^\circ$.
In contrast, the net contribution of multipoles with $\ell \gtrsim 20$ practically
cancels out, producing a plateau of zero correlation if nonuniform weights $w_{\ell}$
are excluded.

Nevertheless, the tail of $C(\theta)$ at $\theta\sim 180^\circ$ is influenced by the
higher-order polynomials, even $\ell \gtrsim 20$. To demonstrate this effect, we have
split the angular correlation function into two pieces:
\begin{eqnarray}\label{eq:CCsplit}
C(\theta)&=&C(\theta\,;\ell \leq \ell_0) +\ C(\theta\,;\ell \geq \ell_0+1)\nonumber \\
&=& \sum_{\ell=2}^{\ell_0}\frac{(2\ell+1)}{4\pi}\ w_{\ell}\ C_{\ell}P_{\ell}(\theta)
+\nonumber \\
&\null&\qquad\sum_{\ell=\ell_0+1}^{\ell_{\rm up}}\frac{(2\ell+1)}{4\pi}\ w_{\ell}\ C_{\ell}P_{\ell}(\theta)\;.
\end{eqnarray}
The first piece, $C(\theta\,;\ell \leq \ell_0)$, corresponds to the $C_{\ell}$ coefficients
that are significantly affected by $u_{\rm min}$ in the integral of Equation~(\ref{eq:Cumin}) and the
different odd-even weighting factors used to produce Figure~3(a). The second summation
runs from $\ell_0+1$ up to $\ell_{\rm up}=401$ (ideally, infinity).

In contrast to conventional wisdom, the contribution of high-$\ell$ multipoles,
$C(\theta\,;\ell \geq \ell_0+1)$, can certainly influence $C(\theta)$ even at angles
$\theta\sim 180^\circ$. To see this, we separately plot in Figure~4(b) the
contributions of even and odd multipoles for $\ell_0>40$. Due to the oscillatory 
behavior of the Legendre polynomials, both contributions add
to always produce a positive correlation at smaller angles, while a delicate balance
exists at $\theta\sim 180^\circ$ that yields zero correlation when the
parity symmetry is exact.  A slight imbalance between odd and even high-$\ell$ multipoles,
however, has a small (but observable) influence on the shape of the tail. We thus conclude
that the value of $C(\theta \simeq 180^\circ)$ contains very interesting information
concerning a possible {\em odd-even parity imbalance even for high-$\ell$ multipoles}.

\section{Discussion}
In spite of the high degree of isotropy in the CMB, certain anomalies have been found that
create some tension with standard inflationary cosmology. The lack of long-range angular
correlations beyond a maximum angle ($\theta_{\rm max} \gtrsim 60^o$) is difficult to
reconcile with the basic inflationary paradigm, which is founded on the principle of a slow-roll
potential producing an accelerated expansion of over 60 e-folds. This maximum correlation
angle instead allows only about 55
e-folds, well below the number required to solve the temperature horizon problem. Moreover,
as shown in Melia \& L\'opez-Corredoira (2017), the feature (i.e., $k_{\rm min}$) 
that maps into such a maximum correlation angle also produces a zero-correlation plateau in the 
two-point angular correlation function at all larger angles.

Meanwhile, it has been known for some time that the angular power spectrum of the
CMB favours a weighting of odd multipoles over even (see, e.g., Kim et al. 2012).
This anomaly could mean a breakdown of the odd-even parity expected in the
cosmological principle, so its study carries great interest. In this paper, we
therefore sought to determine whether these two features, that is, $k_{\rm min}$ and
an odd-even parity imbalance, are related, and/or whether both are required to produce
the best fit to the {\it Planck} 2018 data.

We have incorporated a possible odd-even parity imbalance in our analysis by introducing
nonuniform weighting factors that modify the $C_{\ell}$ coefficients in accordance with the
parity statistic $Q(\ell_{\rm max})$. The outcome of this analysis has resulted in an
excellent fit to the two-point correlation function, including the tail associated with
the odd-parity dominance at $\theta\sim 180^\circ$. We have stressed that this angular
region is influenced not only by the expected low-$\ell$ multipoles, but also
to some degree by the high multipoles when the odd-even parity is broken.

At this stage, we can only speculate about a possible physical origin of $k_{\rm min}$
and/or an odd-even parity imbalance. Certainly, in the context of slow-roll inflation,
the cutoff signals the time at which inflation could have started. This is the
principal reason why a nonzero value of this truncation to the primordial power spectrum
is so restrictive for the ability of inflation to solve the temperature horizon problem
while simultaneously producing the distribution of anisotropies seen in the CMB. A natural
question that arises in this context is whether the presence of a cutoff, $k_{\rm min}$,
should also impact the angular power spectrum itself. The answer appears to be yes,
and it does so in a very intriguing way. It appears that the same value of $u_{\rm min}$
required to optimize the fit to the angular correlation data also completely accounts
for the `missing' power seen in the low-multipole components (Melia et al. 2021).
The fact that the same feature in ${\mathcal{P}}(k)$, that is, a $u_{\rm min}=4.5\pm0.5$
(as we have found in this paper) can account for both empirically derived anomalies
adds weight to its possible reality.

We here demonstrated that both $k_{\rm min}$ and an odd-even parity imbalance
are required to optimize the fit to the {\it Planck} 2018 data, however. So what could be the origin
of this parity violation? There is no known mechanism that can produce such an imbalance
due to inflation on its own. Prior to classicalization, all of the quantum fluctuations
seeded in the early Universe and expanded during inflation were spherically symmetric
(Melia 2021). Suggestions have therefore tended to focus on possible nonstandard
beginnings or trans-Planckian issues. For example, topological models involving multiconnected
universes have been invoked to account for the anomalous cold spot, or the aforementioned
missing power at low multipoles (Efstathiou 2003; Land \& Magueijo 2006). Although these models
can lower the power of the small-$\ell$ multipoles, they apparently cannot create an
asymmetry, however.

At face value, an odd-even parity imbalance might be viewed as a possible trans-Planckian
effect (Brandenberger \& Martin 2013), given that this is the first instance 
following the Big Bang when features measurable today would have emerged into the 
semi-classical Universe (Melia 2020). This topic touches on a broader issue related to 
the self-consistency of basic 
inflationary theory because the quantum fluctuations in the inflaton field would have been seeded
in the so-called Bunch-Davies vacuum (Bunch \& Davies 1978), well below the Planck scale.
It is unclear, however, how or why quantum mechanics as we know it and general relativity could
be used meaningfully to describe the evolution of these fluctuations on scales smaller
than their Compton wavelength (see, e.g., Melia 2020). In other words,
an odd-even parity imbalance may turn out to be a signature of trans-Planckian physics
once a viable theory of quantum gravity is devised, but there is no evidence of this
right now.  Any oscillatory and sharp features in ${\mathcal{P}}(k)$ tend to
become completely smeared out by the time the CMB power spectrum is produced (Bennett et al. 2011).
Other possibilities may also include bouncing cosmologies (Agullo et al. 2020).

If confirmed to be of cosmological origin, an odd-parity dominance would violate the Cosmological
Principle. Together with the missing large-angle correlations, these two features could be
an indication that new physics is required to modify the standard model accordingly, perhaps
even leading to some exciting new discoveries about the origin and evolution
of the Universe.

Regardless of its origin, however, the reality of an odd-over-even parity imbalance 
is becoming more firmly established in the analysis of the {\it Planck} data, making speculation 
such as this interesting to consider.
Nevertheless, the existence of this asymmetry is by no means a certain indication that it
originates from the CMB itself. As we indicated in the introduction, it may simply
be due to an observational artifact. Recently, Creswell \& Naselsky (2021b) discussed 
a link between the parity asymmetry and the low-$\ell$ peak anomaly, established in
the presence of highly asymmetric regions in the sky, due to some foreground contamination
in four regions near the Galactic plane ($[\ell,b]=[212^\circ,21^\circ]$,
$[32^\circ,21^\circ]$, $[332^\circ,8^\circ]$, and $[152^\circ,8^\circ]$).
This asymmetric distribution increases the odd-multipole power, while the deficit
of symmetric regions leads to a corresponding deficit of even-multipole peaks. Therefore,
the odd-even parity imbalance in the CMB could be explained, to a large statistical
significance, as a consequence of an anomalous density of antipodal peaks in the sky
once the dipole contribution from the motion of the Solar System is removed, without
resorting to actual cosmological effects.

Having said that, we can safely conclude that our study in this paper establishes
the compatibility of an infrared cutoff $k_{\rm min}$ in the power spectrum with an odd-over-even
parity imbalance, regardless of its origin, either cosmological or due to contamination, or
an incorrect foreground subtraction. Our analysis has shown that both $k_{\rm min}$ and
the parity asymmetry are necessary in order to provide a best fit to the angular correlation
function of the CMB.

\section{Conclusion}
This brief survey of possible causes of an odd-even parity imbalance is by no means
exhaustive, but it is fair to conclude that a resolution of its origin will probably rely
on new theoretical ideas. This stands in contrast to the meaning of $k_{\rm min}$, which
can indeed in some way be attributed
to the inflaton potential, $V(\phi)$. At least theoretically, there would be no obvious
connection between this cutoff and the odd-even parity imbalance. Nevertheless, the
observational evidence suggests that both are necessary to optimize the fit to the
{\it Planck} 2018 data, as hinted also for the WMAP observations by the earlier work
reported in Kim et al. (2012), although the WMAP measurements had a lower precision
than the {\it Planck} 2018 observations, and these authors did not analyze in
depth the two-point angular correlation function over its whole angular range, as we have
done here for {\it Planck}.

Our main conclusion is that neither the odd-even imbalance nor the cutoff
$k_{\rm min}=4.5/r_d$ on their own and separately are sufficient to minimize
the $\chi^2_{\rm dof}$ of the $C(\theta)$ fit to the {\it Planck} 2018 data. Both
are required, and while a nonzero $k_{\rm min}$ may be attributed to an as yet
undiscovered inflaton potential, the odd-even imbalance would appear to signal
entirely new physics, if it is not simply due to contamination.

\begin{acknowledgements}
This work has been partially supported by Agencia Estatal de Investigaci\'on del Ministerio de Ciencia e Innovaci\'on 
under grant PID2020-113334GB-I00 / AEI / 10.13039/501100011033, by Generalitat
Valenciana under grant PROMETEO/2019/113 (EXPEDITE), by the Center for Research and Development in Mathematics and Applications (CIDMA) through the Portuguese Foundation for Science and Technology (FCT - Funda\c c\~ao para a Ci\^encia e a Tecnologia), references UIDB/04106/2020 and UIDP/04106/2020, by national funds (OE), through FCT, I.P., in the scope of the framework contract foreseen in the numbers 4, 5 and 6
of the article 23, of the Decree-Law 57/2016, of August 29,
changed by Law 57/2017, of July 19 and by the projects PTDC/FIS-OUT/28407/2017,  CERN/FIS-PAR/0027/2019 and PTDC/FIS-AST/3041/2020. This work has further been supported by  the  European  Union's  Horizon  2020  research  and  innovation  (RISE) programme H2020-MSCA-RISE-2017 Grant No.~FunFiCO-777740 and by FCT through Project~No.~UIDB/00099/2020.
\end{acknowledgements}

\begin{small}
\subsection*{References}
Abramowitz, M. (Editor) \& Stegun, I. A.,1970, {\it Handbook of 
Mathematical Functions: with Formulas, Graphs, and Mathematical Tables} (Dover Books on Mathematics, NY)\\
Agullo, I., Kranas, D. \& Sreenath, V., 2021, CQG, 38, 065010 \\
Aluri, P. K. \& Jain, P., 2012, MNRAS, 419, 3378\\
Bennett, C. L. et al., 2003, ApJS, 148, 97\\
Bennett, C. L., Hill, R. S., Hinshaw, G. et al., 2011, ApJ, 192, 17\\
Bouchet, F. R. et al., 2011, {\it COrE (Cosmic Origins Explorer) A White Paper}, (arXiv:1102.2181)\\
Brandenberger, R. H. \& Martin, J., 2013, CQG, 30, 113001\\
Bunch, T. S. \& Davies, P.C.W., 1978, Proc. R. Soc. A, 360, 117\\
Creswell, J. \& Naselsky, P., 2021a, JCAP, 2021, 103\\
Creswell, J. \& Naselsky, P., 2021b, submitted (arXiv:2105.08658)\\
Di Valentino, E., Melchiorri, A. \& Silk, J., 2021, ApJL, 908, L9\\
Dong, F., Yu, Y., Zhang, J., Yang, X. \& Zhang, P., 2021, MNRAS, 500, 3838\\
Efstathiou, G., 2003, MNRAS Lett., 346, L26\\
Errard, J., Feeney, S. M. et al., 2016, JCAP, Issue 03, article id. 052\\
Guth, A. H., 1981, PRD, 23, 347\\
Hanany, S. et al. [NASA PICO], 2019, {\it PICO: Probe of Inflation and Cosmic Origins}, (arXiv:1902.10541)\\
Hinshaw, G. et al., 1996, ApJL, 464, L25\\
Kazanas, D., 1980, ApJL, 241, L59\\
Kim, J., Naselsky, P. \& Hansen, M., 2012, Adv. Astron., 2012, 960509\\
Land, K. \& Magueijo, J., 2006, MNRAS, 367, 1714\\
Linde, A. D., 1982, PLB, 108, 389\\
Liu, J. \& Melia, F., 2020, Proc. R. Soc. A, 476, 20200364\\
L\'opez-Corredoira, M., Sylos Labini, F. \& Betancort-Rijo, J., 2010, A\&A, 513, A3\\
L\'opez-Corredoira, M., 2017, Found. Phys., 47, 711\\
Melia, F., 2014, A\&A, 561, A80\\
Melia, F., 2013a, A\&A, 553, A76\\
Melia, F., 2013b, CQG, 30, 155007\\
Melia, F., 2014, A\&A, 561, A80\\
Melia, F. and L\'opez-Corredoira, M., 2018, A\&A, 610, A87\\
Melia, F., 2018, AJP, 86, 585\\
Melia, F., 2019, EPJ-C, 79, 455\\
Melia, F., 2020, Astron. Nachrichten, 341, 812\\
Melia, F., 2020b, {\it The Cosmic Spacetime} (Taylor \& Francis, Oxford)\\
Melia, F., 2021, PLB, 818, 136632\\
Melia, F., Ma, Q., Wei, J.-J. \& Yu, B., 2021, A\&A, 655, A70\\
Melia, F., 2022, Astron. Nachricten, in press (10.1002/asna.20224010)\\
Mukhanov, V. F., 2005, {\it Physical Foundations of Cosmology}
(Cambridge University Press, Cambridge)\\
Panda, S., Aluri, P. K., Samal, P. K. \& Rath, P. K., 2021, Astropart. Phys., 125, 102493
[erratum: Astropart. Phys., 130, 102582]\\
Peebles, P.J.E., 1980, {\it The Large-Scale Structure of the Universe}
(Princeton University Press, Princeton)\\
Perivolaropoulos, L. \& Skara, F., 2021, (arXiv:2105.05208)\\
Planck Collaboration, Aghanim, N. et al., 2018, A\&A, 641, A86\\
Sachs, R. K. \& Wolfe, A. M., 1967, ApJ, 147, 73\\
Sanchis-Lozano, M. A., Sarkisyan-Grinbaum, E. K.,
Domenech-Garret, J. L. \& Sanchis-Gual, N., 2020, PRD, 102, 035013\\
Starobinsky, A. A., 1979, J. Exp. and Theo. Phys. Lett., 30, 682\\
Schwarz, D. J., Copi, C. J., Huterer, D. \& Starkman, G. D., 2016, CQG, 33, 184001
\end{small}

\begin{appendix}

\section{Expansion factor following decoupling}
The comoving distance to the last scattering surface may be written
\begin{equation}
{r}_d=c\int_{t_d}^{t}\frac{dt'}{a(t')}\;.
\end{equation}
The scale factor $a(t)$ determined from Friedmann's equations for an isotropic and
homogeneous Universe made of matter (dust) and dark energy reads
\begin{eqnarray}
\biggl(\frac{\dot{a}}{a}\biggr)^2&=&H_0^2\ \biggl[\frac{\Omega_m}{a(t)^3}+
\Omega_{\Lambda}\biggr]\nonumber \\
&\rightarrow& a(t) \sim \biggl(\sinh\biggl[\frac{3}{2}\sqrt{\Omega_{\Lambda}}
H_0t\biggr]\biggr)^{2/3}\;,
\end{eqnarray}
where $\Omega_{\Lambda} \simeq 0.7$ and $H_0$ denote the normalized
dark matter density and Hubble parameter today, respectively. To simplify the notation, we define
$\tilde{H}=(3/2)\sqrt{\Omega_{\Lambda}}H_0$, noting that in fact $\tilde{H} \simeq H_0$.

The proper (or physical) distance is given by
\begin{equation}
R_d=\ a(t)\ {r}_d\;,
\end{equation}
so that
\begin{equation}
R_d=\ \sinh^{2/3}{[\tilde{H}t]}\int_{t_d}^{t}\ \frac{dt'}{\sinh^{2/3}[\tilde{H}t']}\;.
\end{equation}
Changing to the variable $x=\sinh[\tilde{H}t]$, we obtain for the indefinite integral
\begin{equation}
\frac{1}{\tilde{H}}\ \int\ \frac{dx}{x^{2/3}(1+x^2)^{1/2}}=\frac{3\sinh^{1/3}{x}}{\tilde{H}}\ _2F_1[1/2,1/6;7/6;-x^2]\;.
\end{equation}
Next, invoking the basic property of the hypergeometric series (Abramowitz \& Stegun 1970),
\begin{eqnarray}
_2F_1[a,b;c;z] &=& (1-z)^{-a}\ _2F_1[a,b-c;c;z/(z-1)]\ \to \nonumber \\
_2F_1[1/2,1/6;7/6;-x^2] &=& \nonumber \\
&\null&\hskip-0.5in\frac{1}{(1+x^2)^{1/2}}\ _2F_1[1/2,1;7/6;x^2/(1+x^2)]\;,
\end{eqnarray}
and reverting back to the original variable, we find the proper distance required
to determine the maximum angle in Equation~(\ref{eq:thetamax}) to be
\begin{eqnarray}\label{eq:rd}
R_d&=&\frac{3\sinh^{2/3}[\tilde{H}t]}{\tilde{H}}\times
\biggl(\frac{\sinh[\tilde{H}t]^{1/3}}{\cosh[\tilde{H}t]}\times\nonumber \\
&\null&\qquad\quad _2F_1\biggl[1/2,1/6;7/6;\tanh^2
[\tilde{H}t]\biggr]- (t \to t_d)\biggr)\;,
\end{eqnarray}
where $t$ here denotes the cosmic time of observation since the Big Bang. If the 
observation is today, then $t=t_0$ is to be identified with the present age of the universe.

\section{Error due to the finite cutoff $\ell_{\rm up}$}
In this appendix, we estimate the accuracy of the Legendre expansion of $C(\pi)$
over $\ell_{\rm up}$ polynomials instead of infinity.  For simplicity, we assume
that the relation $C_{\ell}=2\ell(\ell+1)$ is satisfied for all $\ell$; we therefore obtain
$4\pi\ C(\pi)= 1/4$, as defined in Equation~(\ref{eq:CCC}) for all $\ell$ from 2 to $\infty$.

Thus, we may write
\begin{equation}
4\pi\ C(\pi)= 1/4=\sum_{\ell=2}^{\ell_{\rm up}}\frac{(2\ell+1)}{2\ell(\ell+1)}(-1)^{\ell}\ 
+\ \frac{(-1)^{\ell_{\rm up}+1}}{2(\ell_{\rm up}+1)}\;.
\end{equation}
This equation allows us to estimate the relative error made by neglecting
polynomials of order higher than $\ell_{\rm up}$. In this work we set $\ell_{\rm up}=401,$
so that the relative error is about $0.5\%$. Other effects
such as BAO at large $\ell$ have been neglected throughout this analysis because they
have very little or no influence on the large angles we are focused on in this paper.
\end{appendix}

\end{document}